\documentclass[twocolumn,showkeys,preprintnumbers,amsmath,amssymb]{revtex4}
\usepackage[]{natbib}
\usepackage{graphicx}% Include figure files
\usepackage[latin1]{inputenc}
\usepackage{dcolumn}% Align table columns on decimal point
\usepackage{bm}% bold math
\usepackage{epsfig}
\usepackage{selinput}
\SelectInputMappings{%
  aacute={á},
  ntilde={ñ},
  Euro={€}
}

\begin{document}

\preprint{Submitted for publication}

\title{Emergence of linguistic laws in human voice}% Force line breaks with \\

\author{Iván González Torre$^1$, Bartolo Luque$^{1,2}$, Lucas Lacasa$^{2,*}$, Jordi Luque$^3$ and Antoni Hernández-Fernández$^{4}$}
\email{l.lacasa@qmul.ac.uk, antonio.hernandez@upc.edu}
\affiliation{$^1$Department of Applied Mathematics and Statistics, EIAE, Technical University of Madrid, Plaza Cardenal Cisneros, 28040 Madrid (Spain)\\$^2$School of Mathematical Sciences, Queen Mary University of London, Mile End Road E14NS London (UK)\\$^3$Telefonica Research, Edificio Telefonica-Diagonal 00, Barcelona (Spain)\\$^4$Complexity and Quantitative Linguistics Lab, Laboratory for Relational Algorithmics, Complexity and Learning (LARCA),
Institut de Ci\`encies de l$'$Educaci\'o, 
Universitat Polit\`ecnica de Catalunya, Barcelona (Spain)}%


\begin{abstract}
Linguistic laws constitute one of the quantitative cornerstones of modern cognitive sciences and have been routinely investigated in written corpora, or in the {\it equivalent} transcription of oral corpora. This means that inferences of statistical patterns of language in acoustics are biased by the arbitrary, language-dependent segmentation of the signal, and virtually precludes the possibility of making comparative studies between human voice and other animal communication systems.
%yet both their onset and measurement is currently highly dependent upon the underlying syntax of the communication system: one needs to have an a priori understanding of a particular language and its code to be able to quantify such things as patterns in the organization of words, phrases, or vocabulary. This strong dependence currently limits the observation and study of linguistic laws to written texts, symbolic sequences or to transcriptions of acoustical waves which in the case of animal communication are often biased by the arbitrary or manual segmentation of the signal. 
Here we bridge this gap by proposing a method that allows to measure such patterns in acoustic signals of arbitrary origin, without needs to have access to the language corpus underneath. The method has been applied to six different human languages, recovering successfully some well-known laws of human communication at timescales even below the phoneme and finding yet another link between complexity and criticality in a biological system. 
%This suggests that in human voice -which has recently been shown to operate close to a critical state- linguistic laws emerge naturally, finding a biological system where evolution has linked complexity and criticality. 
These methods further pave the way for new comparative studies in animal communication or the analysis of signals of unknown code.
\end{abstract}

%\pacs{}% PACS, the Physics and Astronomy
\keywords{Zipf $|$ Heaps' law $|$ Brevity law $|$ Communication $|$ Human voice} \maketitle

The main objective of quantitative linguistics is to explore the emergence of statistical patterns (often called linguistic laws) in language and general communication systems (see \cite{quantitativelinguisticshandbook} and \cite{Altmann2016} for a review). The most celebrated of such regularities is 
Zipf's law describing the uneven abundance of word frequencies \cite{Zipf1935psycho, Zipf1949human}. This law presents many variations in human language \cite{i2005variation, baixeries2013evolution, piantadosi2014Zipf, van2015Zipf} but also shows ubiquity \cite{li2002Zipf} in many linguistic scales \cite{ha2002extension}, has been claimed to be universal \cite{corominas2010universality, Ferrer2016compression} and has consequences for syntax and symbolic reference \cite{i2005consequences}. On the other hand Heaps' law, also called Herdan's law \cite{herdan1964quantitative, heaps1978information} states that the vocabulary of a text grows allometrically with the text length \cite{font2013scaling, gerlach2014scaling}, and is mathematically connected with Zipf's law \cite{mandelbrot1961theory, baayen2001word, FontClosCorral2015}, the scaling exponent being dependent on both Zipf law and the vocabulary size. Finally the Zipf's Law of abbreviation (or brevity law for short) is the statistical tendency of more frequent elements in communication systems to be shorter or smaller \cite{Zipf1949human, grzybek2006contributions} and has been recently claimed as an universal trend derived from fundamental principles of information processing \cite{bentz_ferrer2016}. As such, this statistical regularity holds also phonetically \cite{aylett2006language, tomaschek2013word} and implies that the higher the frequency of a word, the the shorter its length or duration, probably caused by a principle of compression \cite{DBLP:journals/corr/Ferrer-i-Cancho15, ferrer2013compression}, although this is a general pattern that can change depending on other acoustical factors like noise \cite{brumm2013animal}, pressure to communicate at long distances calls \cite{ferrer2013failure} or communicative efficiency \cite{Zipf1949human} and energetic constraints \cite{gillooly2010energetic}.\\
\noindent Linguistic laws extend beyond written language and have been shown to hold for different biological data \cite{schwab2014Zipf,piantadosi2014Zipf}. According to some authors \cite{kello2010scaling}, the presence of scaling laws in communication is indicative of the existence of processes taking place across different cognitive scales. Interpreting linguistic laws as scaling laws which emerged in communication systems \cite{FerrerSole2003, NowakKrakauer1999} actually opens the door for speculating on the existence of underlying scale-invariant physical laws operating underneath \cite{Chater1999B17}. Of course, in order to explore the presence or absence of such patterns one needs to directly study the acoustic corpus, i.e. human voice, as every linguistic phenomenon or candidate for language law can be camouflaged or diluted after the transcription process. Notwithstanding the deep relevance of linguistic laws reported in written texts, we still wonder up to which of these laws found in written corpora are related or derive from more fundamental properties of the acoustics of language -and are thus candidates for full-fledged linguistic laws-, or emerge as an artifact of scripture codification.\\

\noindent {\bf Linguistic laws in acoustics? }
Acoustic communication is fully determined by three physical magnitudes extracted from the signals: frequency, energy and time \cite{sueur2006insect, saposhkov1983electroacustica}. The configuration space induced by these magnitudes results from an intrinsic evolutionary relationship between the production and perception of sound systems \cite{McNeilage2012} that further shapes the range of hearing and producing sounds for a variety of life forms \cite{sueur2006insect, Berg_Stork_1995}. 
Since animals use their acoustic abilities both to monitor their environment and to communicate we should expect that natural selection has in some sense optimized these sensorial capabilities, but, despite the great differences that evolution has involved for different animals that communicate acoustically, there are many similarities between their mechanisms of sound production and hearing \cite{Fletcher2014}. Focusing on primates, traditionally language has been distinguished from vocalizations of nonhuman primates only at a qualitative, semantic level \cite{fitch2000}. Interestingly, it is well known that children use statistical cues to segment the input \cite{Saffranetal1996,Kuhletal2008,Romberg_Saffran_2010} and probably share with non-human primates some of these mechanisms, albeit with some differences \cite{Saffranetal2008}. The discovery of these statistical learning abilities has boosted a new approach to the study of language \cite{Kuhl2000, Emberson2016} and suggests that statistical learning of language could be based on patterns or more generally linguistic laws \cite{bentz_ferrer2016}:  research on language acquisition shows that higher frequency facilitates learning, and Zipf's law tell us that vocabulary learning is easier than expected a priori given the skewness of the distribution, for instance. 
 We advance that, as will be shown below, these patterns are already present in the physical sound waves produced by human voice even at levels below those generally considered linguistically significant, i.e., below the phoneme timescale.\\

\begin{figure*}[tbhp]%[tbhp]
\centering
\includegraphics[width=0.65\linewidth]{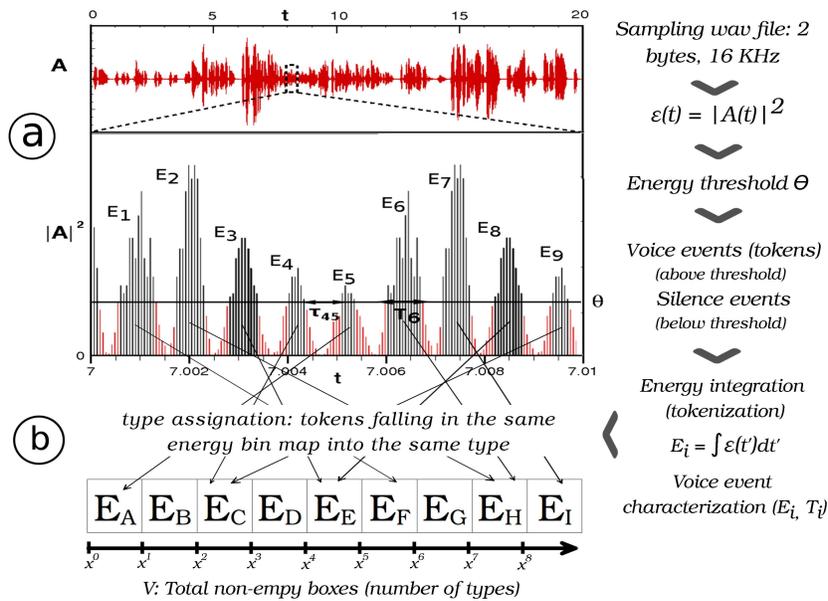}
\caption{This figure illustrates the methodology to extract a sequence of types from an acoustic signal. Waveform series $A(t)$ are sampled at 16KhZ from the system. In (a) we plot the instantaneous energy per unit time $\epsilon (t) = |A|^2(t)$ from an excerpt of the top panel. The energy threshold $\Theta$, defined as the instantaneous energy level for which a fixed percentage of the entire data remains above-threshold, helps us to unambiguously define a token or voice event (a subsequence of time stamps for which $\epsilon (t) > \Theta$) from silence events of duration $\tau$ \cite{luque2015scaling}. The energy released $E$ in a voice event is computed from the integration of the instantaneous energy over the duration of that event  (dark area in the figure denotes the energy released in a given voice event). By performing a logarithmic binning tokens are classified into several bins that we call types. The vocabulary $V$ agglutinates those types that appear at least once.}
\label{fig:Figure1}
\end{figure*}

\noindent Empirical evidence of robust linguistic laws holding in written texts across different human languages has been reported many times (see \cite{Altmann2016, baayen2001word, piantadosi2014Zipf} and references therein), and it has been shown that these laws are not fully observed in random texts \cite{FerrerElvevag2010}. Studies with oral corpus are however much less abundant, and they systematically imply a transcription of the acoustical waves into words in the case of human speech or some ill-defined analog of words in the case of animal communication, as the main segments to analyze \cite{McCowanetal1999, FerrerMcCowan2009}. A few current efforts take a different road and consider other possible written units such as lemmas \cite{Corraletal2015} or compare written and oral production for some linguistic patterns \cite{Nabeshima_Gunji2004, baroni08}, in general showing that frequencies of elements in written corpora can be taken as a rough estimate for their frequency in spoken language \cite{Samlowskietal2011}.

\noindent All in all, the exploration of linguistic laws in oral corpora is scarce. In fact, all linguistic studies in oral and written corpora are influenced by our segmentation decisions and our definition of 'word', intimately biased by an inherently anthropocentric perspective and, of course, by our linguistic tradition. The idea that speech is produced like writing, as a linear sequence of vowels and consonants may indeed be a relic of our scripture technology. As a matter of fact, it is well-known in linguistics that both vowels and consonants are produced linearly but also depend on their surrounding elements: this is the traditional and well-studied concept of coarticulation \cite{Farnetani_Recasens2010}. The boundaries between acoustical elements are therefore difficult to identify if we are not native speakers of a language, and that's yet a crucial problem of phonotactics and speech segmentation and recognition \cite{Glass2003} with differences across languages \cite{TylerCutler2009}.

\noindent Precisely because of this, classical signal segmentation based on the concept of "word" inherited from writing has lead some researchers to search how to transform artificially written corpora into phoneme or syllabic chains with different objectives \cite{Taylor2009}, at the time it involves two major problems in communication studies, namely (i) the impossibility of performing fully objective comparative studies between human and non-human signals \cite{Kuhl2003}, where signals could be physical events, behaviours or structures to which receivers respond \cite{Stegmann2013}. This problem leads researchers sometimes to manually segment acoustic signals guided only by their expertise, and prevents to explore signals of unknown origin including for instance the search for possible extraterrestrial intelligence \cite{Doyleetal2009}. And (ii) a rather arbitrary definition of the units of study guided by ortographic conventions already produces non-negligible epistemological problems at the core of Linguistics  \cite{bunge1984, Kohler2005}.
\\

In this work we explore the acoustic analog of classical linguistic laws in the statistics of speech waveform from extensive real databases \cite{rodriguez2010kalaka} that for the first time extend all the way into the intraphoneme range ($t<10^{-2}$s) \cite{crystal1988segmental}. 
In order to do so in a systematic way, in what follows we will present a methodology that enables the direct linguistic analysis of any acoustical waveform without needs to invoke to {\it ad hoc} codes or assume any concrete formal communication system underneath (Materials and Methods, see figure \ref{fig:Figure1} for an illustration). 
This method only makes use of physical magnitudes of the signal such as energy or time duration which therefore allow for a nonambiguous definition of our units. Speech is indeed a physical phenomenon and as such in this work we interpret it as a concatenation of events of energy release, and propose a mathematically well-defined way for segmenting this orchestrated suite of energy release avalanches \cite{luque2015scaling}.

\noindent We find clear evidence of robust Zipf, Heaps and brevity laws emerging in this context and speculate that this might be due to the fact that human voice seems to be operating close to criticality, hence finding an example of a biological system that, driven by evolution, has linked complexity and criticality.
We expect that this methodology can open a fresh line of research in communication systems where a direct exploration of underlying statistical patterns in acoustic signals is possible without needs to predetermine any of the aforementioned non-physical concepts, and hope that this will allow researchers to develop comparative studies between human language and other acoustical communication systems or even to unravel whether if a generic signal shares these patterns \cite{Altmann2016}.\\

\noindent {\bf Data. }
For this work we have used a TV broadcast speech database named KALAKA-2 \cite{rodriguez2010kalaka}. Originally designed for language recognition evaluation purposes, it consists of wide-band TV broadcast speech recordings (4h per language sampled using 2 bytes at a rate of 16kHz) ranging six different languages: Basque, Catalan, Galician, Spanish, Portuguese and English and encompassing both planned and spontaneous speech throughout diverse environmental conditions, such as studio or outside journalist reports but excluding telephonic channels.\\

\noindent {\bf The method. }
\noindent The objects under study, speech waveforms or otherwise any generic acoustic signal, are fully described by an amplitude time series $A(t)$ (see figure \ref{fig:Figure1} for an illustration of the method). In order to unambiguously extract a sequence of symbols -the equivalent to words and phrases- from such signal without the need to perform \textit{ad hoc} segmentation, we start by considering the semi-definite positive magnitude $\epsilon(t) = |A(t)|^{2}$ which, dropping irrelevant constants, has physical units of energy per time (SI for additional details on speech waveform statistics). By defining an energy threshold $\Theta$ in this signal \cite{luque2015scaling} we will unambiguously separate voice events (above threshold) from silence events (below threshold).
More concretely, $\Theta$ is defined as a relative percentage and its actual value in energy units depends on the signal variability range: for example $\Theta = 80\%$ means that $20\%$ of the data falls under this energy level. It has been shown that $\Theta$ decimates the signal similarly to a real-space Renormalization Group (RG) transformation \cite{luque2015scaling,CorralJSTAT}, in such a way that increasing $\Theta$ induces a flow in RG space. Systems operating close to a critical state lie in an unstable fixed point of this RG flow and its associated signal statistics are therefore shown to be threshold-invariant. 
Now, $\Theta$ not only works as an energy threshold that filters out background or environmental noise (noise filtering being a key aspect that species have learned to perform \cite{brumm2005acoustic, brumm2013animal}) but, as previously stated, enable us to unambiguously define what we call a {\it token} or voice event, that is, a sequence of consecutive measurements $\epsilon(t)>\Theta$, from a silence event of duration $\tau$. Each token is in turn characterized by a duple $(E_v,T_v)$ where $T_v$ is the duration of the event and $E_v$ corresponds to the total energy released during that event $E_v=\int_t^{t+T_v}\epsilon(t')dt'$ obtained summing up the instantaneous energy over the duration of the event. Accordingly, the signal is effectively transformed to an ordered sequence of tokens $\{E_v(i), T_v(i)\}$, each of these being separated by silence events of highly heterogeneous durations $\tau$ which, incidentally, are known to be power law distributed \cite{luque2015scaling}. Finally, by logarithmically binning the scale of integrated energies we can assign an energy label (the bin) to each token, hence mapping the initial acoustic signal into a symbolic sequence of fundamental units which we call {\it types}. The logarithmic binning is justified here invoking Fechner-Weber's law \cite{Chater1999B17}.
Note that two tokens whose integrated energy fall in the same energy range are mapped to the same type even if their duration can be different, so in principle several tokens could map into the same type (see SI for a table of Type/Token ratios). By default we define as many bins as voice events such that the set of bins can be understood as an abstraction of a universal language vocabulary, accordingly some bins might be empty and in general each bin will occur with uneven frequencies. As such, types can be understood as acoustically-based universal abstractions of a fundamental unit, an abstract version of words or phonemes that appear intertwined in a signal with characteristic patterns.\\ %Incidentally, a hierarchy of different semantic levels could then be defined by setting lower and upper bounds to the silence between tokens; for instance tokens separated by silences of duration larger than $\tau_{\text{lb}}$ but shorter than $\tau_{\text{ub}}$ define a higher semantic category called a {\it type}.\\

\noindent To summarize, with this methodology we are able to map an arbitrary acoustic signal into a sequence of types separated by silence events (figure 1).
%where different semantic structures are obtained as subsequences. 
Standard linguistic laws can then be directly explored in acoustic signals without needs to have an \textit{a priori} knowledge neither of the signal code nor of the adequate segmentation process or the particular syntax of the language underlying the signal. This protocol is thus independent of the communication system and can be used to make unbiased comparisons across different systems and signals. %bridging the gap in the state of the art that assumes that the modeling of a semiotic system imposes the need of splitting certain axioms that are necessary to carry out solid research  \cite{kohler1990elemente}. 
Needless to say, results could in principle depend on the particular value of $\Theta$, as this scans the signals at different energy thresholds. However human voice has been recently shown to be invariant under changes in $\Theta$ -an evidence of self-organized criticality (SOC) \cite{Bak} in this system- and, accordingly, parameter-free laws can be extracted using a proper collapse theory as it will be shown in the results section. Finally, in order to guarantee that the emergence of linguistic laws is only due to the structure and correlations of the signal and not due to the process of symbolization we will compare the results obtained from speech signals to properly defined null models which randomize the signal $\epsilon(t)$. These null models thus maintain the marginal instantaneous energy distribution and remove any other correlation structure, yielding non-Gaussian white noise with a fat-tailed marginal distribution.\\

%\showmatmethods % Display the Materials and Methods section

\section*{RESULTS}

\begin{figure}%[tbhp]
\centering
\includegraphics[width=1\linewidth]{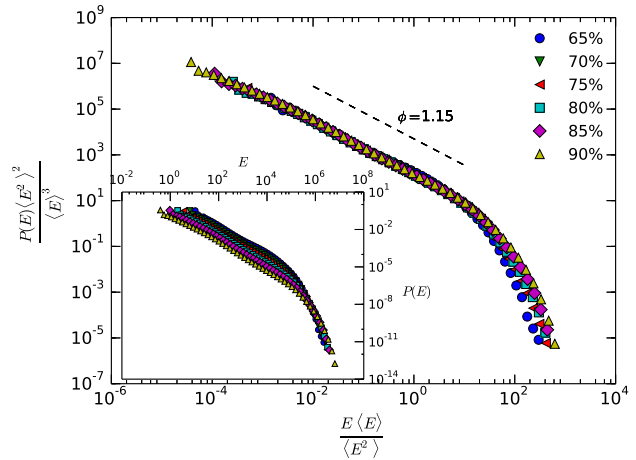}
\caption{Log-log plot of the collapsed shape (see the text) and threshold-independent energy release distribution $P(E)$ in the case of Spanish language for several thresholds, after logarithmic binning. (Inset panel) Non-collapsed distribution, for different thresholds.}
\label{fig:Figure_colapso_E_integrada}
\end{figure}
\noindent {\bf Gutenberg-Richter law. }
The energy $E$ released during voice events is a direct measure of the vocal fold response function under air pressure perturbations, and its distribution $P_{\Theta}(E)$ could in principle depend both on the threshold $\Theta$ and on the language under study. In the inset of figure \ref{fig:Figure_colapso_E_integrada} we observe that $P_{\Theta}(E)$ is power law distributed over about six decases, saturated by an exponential cut-off. This distribution has been interpreted before as the analogue of a Gutenberg-Richter law in voice, as the precise shape of energy release fluctuations during voice production parallel those occurring in earthquakes \cite{luque2015scaling}. As increasing $\Theta$ induces a flow in RG space, systems which lie close to a critical point (unstable fixed point in RG space) show scale invariance under $\Theta$ and hence the distributions can be collapsed into a $\Theta$-independent shape, thereby eliminating the trivial dependence on $\Theta$. This has been shown to be the case for human voice and accordingly (technical details can be found in the SI) we can express the collapsed energy distribution as $P(E) = E^{-\phi} {\cal F}(E/E_{\xi}) $ for $E>>E_{l}$, where $E_{l}$ is the lower limit beyond which this law is fulfilled, $\cal F$ is a scaling function and the relevant variable is $\phi$, the scaling exponent. In order to collapse every curve $P_{\Theta}(E)$ the theory predicts to rescale $E\to E\langle E\rangle /\langle E^2\rangle, \ P_{\Theta}(E)\to P_{\Theta}(E)\langle E^2\rangle^2/\langle E\rangle^3$.
In the outset of Figure \ref{fig:Figure_colapso_E_integrada} we show the result of this analysis for the case of Spanish language, where we find that $\phi\approx 1.15$, this exponent being approximately language-independent (see table \ref{table:tablas_slopes} for other languages and SI for additional details). Interestingly, these exponents are compatible with those found in rainfall, another natural system that has been shown to be compatible with SOC dynamics \cite{rainfall}, and cannot be explained by simple null models \cite{luque2015scaling}. In what follows we explore the emergence of classical linguistic laws in these acoustic signals.\\

\begin{table}
\begin{center}
\begin{tabular}{|c|c|c|c|c|}\hline
Exponent		&$\phi$  & $\zeta$ &$\alpha$&$\beta$	\\\hline
%Exponent		& $\phi$ & $\zeta$ & $\alpha$ & $\beta$	\\\hline
Basque			& $1.13 \pm 0.04 $ & $1.77 \pm 0.14 $ & $0.90 \pm 0.03$ & $3.1 \pm 0.3$  \\\hline
Catalan 		& $1.17 \pm 0.05$ & $1.89 \pm 0.14$ & $0.92 \pm 0.03$& $2.8 \pm 0.4$ 	\\\hline
English 		& $1.16 \pm 0.05$ & $1.85 \pm 0.14$ & $0.91 \pm  0.01$& $2.9 \pm 0.3$	\\\hline
Galician 		& $1.18 \pm 0.04$ & $1.80 \pm 0.14$ & $0.89 \pm 0.03$& $2.9 \pm 0.4$ 	\\\hline
Portuguese 		& $1.16 \pm 0.05$ & $1.77 \pm 0.14$ & $0.91 \pm 0.01$& $3.0 \pm 0.3$	\\\hline
Spanish 		& $1.15 \pm 0.04$ & $1.79 \pm 0.14$ & $0.91 \pm 0.03$& $2.8 \pm 0.4$ 	\\\hline
\end{tabular}
\end{center}
\caption{Summary of scaling exponents associated to the energy release distribution ($\phi$), Zipf's law ($\zeta$), Heaps' law ($\alpha$) and Brevity law ($\beta$) for the six different languages. Power law fits are performed using maximum likelihood estimation (MLE) following Clauset \cite{clauset2009power} and goodness-of-fit test and confidence interval are based on Kolmogorov-Smirnov (KS) tests. In all cases, KS are greater than 0.99. Exponents associated to energy release are compatible with those found in rainfall \cite{rainfall}. Results are compatible with the hypothesis of language-independence. }
\label{table:tablas_slopes}
\end{table}

\noindent {\bf Zipf's Law. }
\begin{figure}%[tbhp]
\centering
\includegraphics[width=1\linewidth]{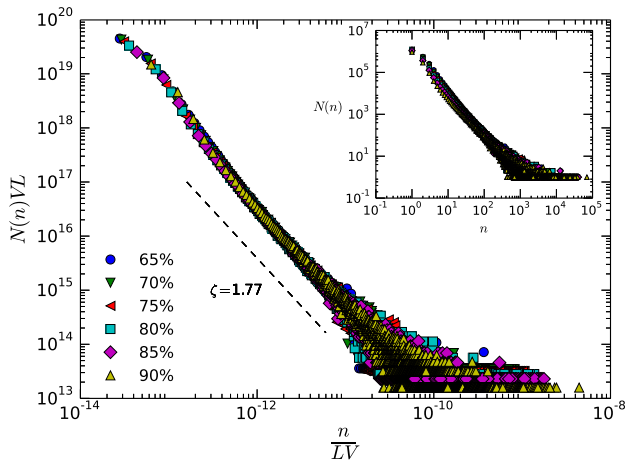}
\caption{Log-log plot of Zipf's law for the case of Basque language (SI for equivalent results in other languages), for different thresholds $\Theta$. The inset panel shows the raw, threshold-dependent distributions and the outer panel Zipf's law has been collapsed (see the text for details).}
\label{fig:Figure_Zipf}
\end{figure}
The illustrious George Kingsley Zipf formulated a statistical observation which is popularly known as Zipf's law \cite{Morenoetal2016}.
In its original formulation \cite{Zipf1935psycho}, it establishes that in a sizable sample of language the number of different words (vocabulary) ${\cal N}(n)$ which occur exactly $n$ times decays as ${\cal N}(n) \sim n^{-\zeta}$, where the exponent $\zeta$ varies from text to text \cite{zanette2014statistical} but is usually close to $2$. An alternative and perhaps more common formulation of this law \cite{Zipf1949human} is defined in terms of the rank, such that if words are ranked in decreasing order by their frequency of appearance then the number of occurrences of words with a given rank $r$ goes like $n(r)\sim r^{-z}$, where
it is easy to see that both exponents are related via $z = \frac{1}{\zeta - 1}$ and thus $z$ approximates to 1 \cite{font2013scaling,5395374}. Here for convenience we make use of the former and explore ${\cal N}_{\Theta}(n)$ applied to the statistics of types. Again ${\cal N}_{\Theta}(n)$ could in principle depend on the threshold but assuming that the signal complies with the scale-invariance mentioned above, one can collapse all threshold-dependent curves into a universal shape and thus remove any dependence on this parameter by rescaling $n\to n/LV, \ {\cal N}_{\Theta}(n)\to {\cal N}_{\Theta}(n)VL$ where $V$ is the total number of different types present in the signal and $L$ as the total number of tokens (see SI for technical details). Results are shown in the case of Basque language in figure \ref{fig:Figure_Zipf}, where a clear threshold-independent decaying power law emerges with a scaling exponent $\zeta \approx 1.77$. Analogous results with compatible exponents for other languages can be found in SI and table \ref{table:tablas_slopes}. Null models systematically deviate from these results, and neither display the characteristic power law decay nor any invariance under variation of the energy threshold (SI).\\

\noindent {\bf Heaps' law. }
\begin{figure}[t]
\centering
\includegraphics[width=1\linewidth]{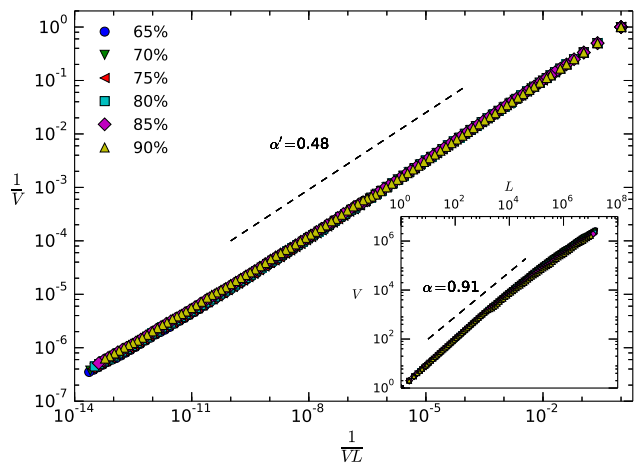}
\caption{Log-log plot of the Heaps' law for the Portuguese sample and several thresholds. In the inner panel we show how the number of different tokens (V) increases sublinearly with the size of the series (L), where the slope can be estimated properly for about three decades. In the main panel we show the collapses curves following \cite{font2013scaling}, where the new scaling exponent $\alpha'\approx 0.48$ is related with the original (see the text) and leads to $\alpha\approx 0.91$. Results for other languages are found in table \ref{table:tablas_slopes}.}
\label{fig:Figure_heap}
\end{figure}
Together with Zipf's law and connected mathematically (see \cite{Altmann2016} and references therein), the second classical linguistic law is Heaps' law, the sublinear growth of the number of different words $V$ in a text with text size $L$ (measured in total number of words): $V \sim L^{\alpha}, \ \alpha <1$ \cite{herdan1964quantitative, egghe2007untangling} (a constant rate for appearance of new words leads to $\alpha=1$). Here the vocabulary $V$ is defined as the total number of different types that appear in the signal, whereas $L$ is defined as the total number of tokens found for a given threshold. Results are shown for a specific language in figure \ref{fig:Figure_heap} (see SI for the rest). In the outset panel we present the collapsed, threshold-independent curves, where again we find a scaling law with an effective exponent $\alpha'$ related to the original exponent $\alpha'= \alpha/(1+\alpha)$. In this case equivalent computation on the null model yield a Heaps law with the trivial exponent $\alpha\approx 1$ (SI).\\

\noindent These results are quantitatively consistent with previous results on written texts \cite{font2013scaling}. In particular, several authors \cite{mandelbrot1961theory} point out that, at least asymptotically, the relation $\zeta = 1+ \alpha$ holds with good approximation, and this is on reasonably good agreement with our findings in human voice as well.
Interestingly, a recent work \cite{lu2013deviation} has found that, as opposed to Indoeuropean (alphabetically based) languages, Zipf's law breaks down and Heaps' law reduces to the trivial case for written texts in Chinese, Japanese, Korean and other logosyllabic languages. Applying our methodology in a database of logosyllabic languages could thus evaluate to which extent those differences arise also in human voice.\\
%that showed the robustness of Zipf's law under changes in system size and a generalization of Heaps' law derived of the stability of the frequency distribution that allows to predict the growth of vocabulary size with text length. 

%For the case of speech events of human voice there are two interpretations for the growth of the text. In the first one, it is compared the different speech events (vocabulary) that appear with the total speech events (Figures and ). The growth of the vocabulary is sub-linear being almost the same for all languages. Differences with the null model... In the second interpretation we can consider the growth of the vocabulary as a function of time ( and )...

\begin{figure}[t]
\centering
\includegraphics[width=1\linewidth]{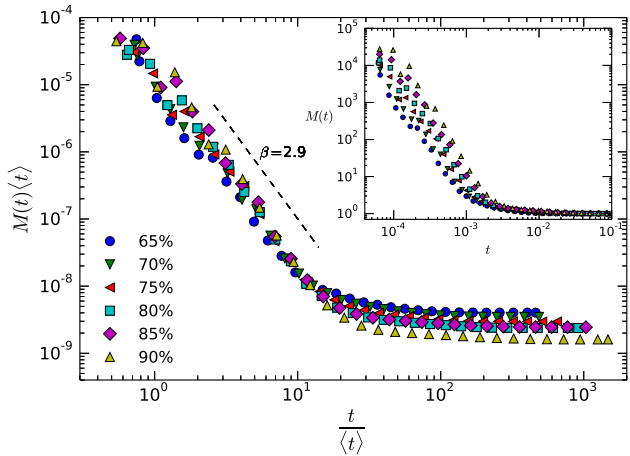}
\caption{Log-log plot of the Brevity law in the case of English, for several thresholds. In the inner panel we plot, for different thresholds, the histogram $M_{\Theta}(t)$ that describes the relative frequency of a type of mean duration $t$. In every case we find a monotonically decreasing curve which yields a brevity law.
In the outset panel we present the collapsed, threshold-independent curve $M(t)$, that evidences an initial power law decay with an exponent $\beta\approx 2.9$. }
\label{fig:Figure_brevity}
\end{figure}

\noindent {\bf Brevity Law. }
The tendency of more frequent words to be shorter \cite{Zipf1935psycho, Zipf1949human, grzybek2006contributions} can be generalized as the tendency of more frequent elements to be shorter or smaller, and its origin has been suggested to be related to optimization and information compression arguments in connection with other linguistic laws \cite{Gustisonetal2016}. In acoustics, spontaneous speech indeed tends to obey this law after text segmentation \cite{greenbergetal2003}, and has been found also in other non-human primates \cite{ferrer2013failure, Gustisonetal2016}. Here we can test brevity law in essentially two different ways. First, note that voice events (tokens) map into types according to the logarithmic binning of their associated energy, hence voice events with different duration might yield the same type as previously noted. Thus for each type we can compute its mean duration $t$ averaging over all voice events that fall within that type, and then plot the histogram $M_{\Theta}(t)$ that describes frequency of each type versus its mean duration. Brevity law would require $M_{\Theta}(t)$ to be a monotonically decreasing function. These results are shown for a particular language in log-log scales in figure \ref{fig:Figure_brevity}, finding initially a power law decaying relation which is indicative of a brevity law (results are again found to be language independent, see SI for additional results). The inset provide the threshold-dependent distributions and the outset panel provides the collapsed, threshold-independent shape $M(t)$ (see table \ref{table:tablas_slopes} for scaling exponents). Again in this case results in null models deviate from such behavior (SI) and are clearly different from the random typing \cite{Ferrer2016compression}. Alternatively, one can also directly observe the duration frequency at the level of voice events, finding similar results (see SI). 
%The usual explanation given to this law is based on the principle of compression (the information-theoretic principle of minimizing the expected length of a code) and maximum efficiency in the theory of information (Gustison et al, 2016). Thus the elements most used in the language tend to be shorter in order to use less energy and time to produce it (Zipf, 1945).

\section*{DISCUSSION}
In this work we have explored the equivalent of linguistic laws directly in acoustic signals. We have found that human voice -which actually complies to SOC dynamics with critical exponents compatible with those found in rainfall \cite{rainfall}- manifests the analog of classical linguistic laws found in written texts (Zipf's law, Heaps' law and the brevity law or law of abbreviation). These laws are found to be invariant under variation of the energy threshold $\Theta$, and can be collapsed under universal functions accordingly. As $\Theta$ is the only free parameter of the method, this invariance determines that the results are not afflicted by ambiguities associated to arbitrarily defining unit boundaries.
Results appear to be robust across languages and timescales (ranging six different Indoeuropean languages and different scales extending all the way into the intraphoneme range, and invariant under energy threshold variation). Interestingly, an equivalent analysis performed on null models defined by randomizing the signal $\epsilon(t)$ (yielding white noise with the same instantaneous energy distribution of the original signal) fail to reproduce this phenomenology (SI). 
The concrete range of exponents found for both Zipf and Heaps laws are compatible between each other and somewhat similar -but not identical- to the typical ones observed in the literature for written texts \cite{Altmann2016, piantadosi2014Zipf, Morenoetal2016,lu2013deviation}, whereas to the best of our knowledge this is the first observation of scaling behavior with a clear exponent in the case of brevity law in speech. Actually, our finding of a power law in brevity law differs from the case of random typing where a power law doesn't conform \cite{Ferrer2016compression}.\\

\noindent The specific and complex alternation of air stops (silences) intertwined with voice production are at the core of the microscopic voice fluctuations. During voice production, acoustic communication is governed by the so-called biphasic cycle (breath and glottal cycle, see \cite{macNeilage08book} for a review) that together with some other acoustic considerations (pitch period, voice onset time, the relation between duration, stress and syllabic structure \cite{greenbergetal2003}) determines the microscopic structure of human voice, including silence stops. However, these timescales are in general very large: as previously stated, this current study focus and scans voice properties even at intraphonemic timescales, where the statistical laws of language emerge directly from the physical magnitudes that govern acoustic communication.
Our results therefore open the possibility of speculating whether the fact that these laws have been found in upper levels of human communication might be a result of a scaling process and a byproduct of the physics rather than derived from the choice of the typical units of study on the analysis of written corpus (phonemes, syllabus, words,...), like differences between analysis of Indoeuropean and logosyllabic languages demonstrates \cite{lu2013deviation}. 
As a matter of fact, in a previous work human voice has been framed within self-organized criticality (SOC), speculating that the fractal structure of lungs drives human voice close to a critical state \cite{luque2015scaling}, this mechanism being ultimately responsible for the microscopic self-similar fluctuations of the signal. This constitutes a new example of the emergence of SOC in a physiological system, different in principle from the classical one found in neuronal activity \cite{Kello2013}.
One could thus speculate that the emergence of these complex patterns is just a consequence of the presence of SOC, what in turn would support the physical origin of linguistic laws. From an evolutionary viewpoint, under this latter perspective human voice, understood as a communication system which has been optimized under evolutionary pressures, would constitute an example where complexity (described in terms of robust linguistic laws) emerges when a system is driven close to criticality, something reminiscent of the celebrated edge of chaos hypothesis \cite{langton1990computation}.\\

\noindent More generally, the method used and proposed here also addresses the longstansing problem of signal segmentation. It has been acknowledged that there is no such thing as a 'correct' segmentation strategy \cite{kvale1993segmentation}. In written corpus white space is usually taken as an easy marker of the separation between words, however this is far from evident in continuous speech where separation between words or concepts is technologically harder to detect, conceptually vague and probably ill-defined. Few exceptions that used oral corpus for animal communication still require to define ad hoc segmentation algorithms \cite{ferrer2013failure}, or manual segmentation strategies which usually give an arbitrary or overestimated segmenting times \cite{Gustisonetal2016}, what might even raise epistemological questions. As such, this segmentation problem unfortunately has prevented wider, comparative studies in areas such as animal communication or the search for signs of possible extraterrestrial intelligence from radio signals (in this line only few proposals have been made \cite{Doyleetal2009}). By varying the energy threshold the method presented here automatically partitions and symbolizes the signal at various energy scales, providing a recipe to establish an automatic, general and systematic way of segmenting and thus enabling comparison of across acoustic signals of arbitrary origion for which we may lack the syntax, code or exact size of its constituents.\\

\noindent To round off, we hope that this work paves the way for new research avenues in comparative studies. Open questions that deserve further work abound; just to name a few: in the light of this new method, what can we say about the acoustic structure in other animal communication systems? Can we find evidence of universal traits in communication that do not depend on a particular species but are only physically and physiologically constrained, or on the other hand are linguistic universals a myth \cite{evans2009myth-gsc}? How these  laws evolve with aging \cite{baixeries2013evolution}? Are they affected by cognitive or pulmonary diseases? What is the precise relation between SOC and linguistic laws in this context? And in particular, can we find mathematical evidence of a minimal, analytically tractable SOC model that produce these patterns? These and other questions are interesting avenues for future work.

\acknowledgments{We acknowledge financial support from ONCE foundation (IGT), grant 2014SGR890 (MACDA) from AGAUR, Generalitat de Catalunya and the APCOM Project TIN2014-57226-P (AHF), grant FIS2013-41057P from Ministerio de Economia y Competitividad (BL and IGT). BL acknowledges the hospitality and support of Queen Mary University of London, where part of this research was developed, and a Salvador de Madariaga fellowship. 
%We are also grateful to R.Ferrer-i-Cancho and A.Corral for helpful comments and valuable discussions.
}

%\showacknow % Display the acknowledgments section

% Bibliography

% \bibliographystyle{unsrt}
% \bibliography{bibliography}

\begin{thebibliography}{84}

\bibitem{quantitativelinguisticshandbook} Kohler, R., Altmann, G., and Piotrowski, R. (2008). Quantitative Linguistics. De Gruyter Mouton, Boston

\bibitem{Altmann2016} Altmann, E. G. and Gerlach, M. (2016). Statistical Laws in Linguistics, in Creativity and Universality in Language, pages 7-26. Springer, Cham

\bibitem{Zipf1935psycho} Zipf, G. K. (1935). The psycho-biology of language

\bibitem{Zipf1949human} Zipf, G. K. (1949). Human behavior and the principle of least effort. Addison-Wesley press

\bibitem{i2005variation} i Cancho, R. F. (2005). The variation of Zipf's law in human language. The European Physical Journal B-Condensed Matter and Complex Systems, 44(2):249-257

\bibitem{baixeries2013evolution} Baixeries, J., Elvevåg, B., and Ferrer-i Cancho, R. (2013). The evolution of the exponent of Zipf's law in language ontogeny. PloS one, 8(3):e53227

\bibitem{piantadosi2014Zipf} Piantadosi, S. T. (2014). Zipf's word frequency law in natural language: A critical review and future directions. Psychonomic Bulletin \& Review, 21(5):1112-1130

\bibitem{van2015Zipf} van Egmond, M., van Ewijk, L., and Avrutin, S. (2015). Zipf's law in non-fluent aphasia. Journal of Quantitative Linguistics, 22(3):233-249

\bibitem{li2002Zipf} Li, W. (2002). Zipf's law everywhere. Glottometrics, 5:14-21

\bibitem{ha2002extension} Ha, L. Q., Sicilia-Garcia, E. I., Ming, J., and Smith, F. J. (2002). Extension of Zipf's law to words and phrases. In Proceedings of the 19th international conference on Computational linguistics-Volume 1, pages 1-6. Association for Computational Linguistics

\bibitem{corominas2010universality} Corominas-Murtra, B. and Solé, R. V. (2010). Universality of Zipf's law. Physical Review E, 82(1):011102

\bibitem{Ferrer2016compression} Ferrer i Cancho, R. (2016b). Compression and the origins of Zipf's law for word frequencies. Complexity

%\bibitem{Ferrer-i-Cancho16arxiv}  Ferrer i Cancho, R. (2016a). Compression and the origins of Zipf's law for word frequencies. CoRR, abs/1605.01326 

\bibitem{i2005consequences} Ferrer i Cancho, R., Riordan, O., and Bollobás, B. (2005). The consequences of Zipf's law for syntax and symbolic reference. Proceedings of the Royal Society of London B: Biological Sciences, 272(1562):561-565

\bibitem{herdan1964quantitative} Herdan, G. (1964). Quantitative linguistics. Butterworth

\bibitem{heaps1978information} Heaps, H. S. (1978). Information retrieval: Computational and theoretical aspects. Academic Press, Inc.

\bibitem{font2013scaling} Font-Clos, F., Boleda, G., and Corral, Á. (2013). A scaling law beyond Zipf's law and its relation to heaps' law. New Journal of Physics, 15(9):093033

\bibitem{gerlach2014scaling} Gerlach, M. and Altmann, E. G. (2014). Scaling laws and fluctuations in the statistics of word frequencies. New Journal of Physics, 16(11):113010

\bibitem{mandelbrot1961theory} Mandelbrot, B. (1961). On the theory of word frequencies and on related markovian models of discourse. Structure of language and its mathematical aspects, 12:190-219

\bibitem{baayen2001word} Baayen, R. H. (2001). Word frequency distributions, volume 18. Springer Science \& Business Media

\bibitem{FontClosCorral2015} Font-Clos, F. and Corral, A. (2015). Log-log convexity of type-token growth in Zipf's systems. Phys. Rev. Lett., 114:238701 

\bibitem{grzybek2006contributions} Grzybek, P. (2006). Contributions to the science of text and language: word length studies and related issues, volume 31. Springer Science \& Business Media

\bibitem{bentz_ferrer2016} Bentz, C. and Ferrer i Cancho, R. (2015). Zipf's law of abbreviation as a language universal. Lorentz Center Workshop, Leiden

\bibitem{aylett2006language} Aylett, M. and Turk, A. (2006). Language redundancy predicts syllabic duration and the spectral characteristics of vocalic syllable nuclei. The Journal of the Acoustical Society of America, 119(5):3048-3058

\bibitem{tomaschek2013word} Tomaschek, F., Wieling, M., Arnold, D., and Baayen, R. H. (2013). Word frequency, vowel length and vowel quality in speech production: an ema study of the importance of experience. In Interspeech, pages 1302-1306

\bibitem{DBLP:journals/corr/Ferrer-i-Cancho15} i Cancho, R. F., Bentz, C., and Seguin, C. (2015). Compression and the origins of Zipf's law of abbreviation. arXiv:1504.0488

\bibitem{ferrer2013compression} Ferrer-i Cancho, R., Hernández-Fernández, A., Lusseau, D., Agoramoorthy, G., Hsu, M. J., and Semple, S. (2013). Compression as a universal principle of animal behavior. Cognitive science, 37(8):1565-1578

\bibitem{brumm2013animal} Brumm, H. (2013). Animal communication and noise. Springer

\bibitem{ferrer2013failure} Ferrer i Cancho, R. and Hernández-Fernández, A. (2013). The failure of the law of brevity in two new world primates. statistical caveats. Glottotheory: International Journal of Theoretical Linguistics, 4(1):45-55

\bibitem{gillooly2010energetic} Gillooly, J. F. and Ophir, A. G. (2010). The energetic basis of acoustic communication. Proceedings of the Royal Society of London B: Biological Sciences, 277(1686):1325-1331

\bibitem{piantadosi2011word} Piantadosi, S. T., Tily, H., and Gibson, E. (2011). Word lengths are optimized for efficient communication. Proceedings of the National Academy of Sciences of the USA, 108(9):3526-3529

\bibitem{schwab2014Zipf} Schwab, D. J., Nemenman, I., and Mehta, P. (2014). Zipf's law and criticality in multivariate data without fine-tuning. Physical Review Letters, 113(6):068102

\bibitem{kello2010scaling} Kello, C. T., Brown, G. D., Ferrer-i Cancho, R., Holden, J. G., Linkenkaer-Hansen, K., Rhodes, T., and Van Orden, G. C. (2010). Scaling laws in cognitive sciences. Trends in cognitive sciences, 14(5):223-232

\bibitem{FerrerSole2003}  Ferrer i Cancho, R. and Solé, R. V. (2003). Least effort and the origins of scaling in human language. Proceedings of the National Academy of Sciences of the USA, 100(3):788-791

\bibitem{NowakKrakauer1999} Nowak, M. A. and Krakauer, D. C. (1999). The evolution of language. Proceedings of the National Academy of Sciences of the USA, 96(14):8028-8033

\bibitem{Chater1999B17}  Chater, N. and Brown, G. D. (1999). Scale-invariance as a unifying psychological principle. Cognition, 69(3):B17-B24

\bibitem{sueur2006insect} Sueur, J. and Drosopoulos, S. (2006). Insect species and their songs. Taylor and Francis, Boca Raton

\bibitem{saposhkov1983electroacustica} Saposhkov, M. A. (1983). Electroacustica. Reverte

\bibitem{McNeilage2012} MacNeilage, P. F. (2011). The evolution of phonology. Oxford University Press

\bibitem{Berg_Stork_1995} Berg, R. and Stork, D. (1995). The Physics of Sound. Prentice Hall

\bibitem{Fletcher2014} Fletcher, N. H. (2014). Animal bioacoustics. In Springer Handbook of Acoustics, pages 821-841. Springer

\bibitem{fitch2000} Fitch, W. T. (2000). The evolution of speech: a comparative review. Trends in Cognitive Sciences, 4(7):258-267

%\bibitem{Saffranetal2008} Saffran, J. R., Hauser, M., Seibel, R., Kapfhamer, J., Tsao, F., and Cushman, F. (2008). Grammatical pattern learning by human infants and cotton-top tamarin monkeys. Cognition 107, 479–500.

\bibitem{Saffranetal1996} Saffran, J. R., Aslin, R. N., and Newport, E. L. (1996). Statistical learning by 8-month-old infants. Science, 274(5294):1926-1928

\bibitem{Kuhletal2008} Kuhl, P. K., Conboy, B. T., Coffey-Corina, S., Padden, D., Rivera-Gaxiola, M., and Nelson, T. (2008). Phonetic learning as a pathway to language: new data and native language magnet theory expanded (nlm-e). Philosophical Transactions of the Royal Society of London B: Biological Sciences, 363(1493):979-1000

\bibitem{Romberg_Saffran_2010} Romberg, A. R. and Saffran, J. R. (2010). Statistical learning and language acquisition. Wiley Interdisciplinary Reviews: Cognitive Science, 1(6):906-914

\bibitem{Saffranetal2008} Saffran, J., Hauser, M., Seibel, R., Kapfhamer, J., Tsao, F., and Cushman, F. (2008). Grammatical pattern learning by human infants and cotton-top tamarin monkeys. Cognition, 107(2):479-500

\bibitem{Kuhl2000} Kuhl, P. K. (2000). A new view of language acquisition. Proceedings of the National Academy of Sciences, 97(22):11850-11857

\bibitem{Emberson2016} Emberson, L. L. and Rubinstein, D. Y. (2016). Statistical learning is constrained to less abstract patterns in complex sensory input (but not the least). Cognition, 153:63-78

\bibitem{FerrerElvevag2010} Ferrer-i Cancho, R. and Elvevåg, B. (2010). Random texts do not exhibit the real Zipf's law-like rank distribution. PLoS One, 5(3):e9411

\bibitem{McCowanetal1999} McCowan, B., Hanser, S. F., and Doyle, L. R. (1999). Quantitative tools for comparing animal communication systems: information theory applied to bottlenose dolphin whistle repertoires. Animal behaviour, 57(2)

\bibitem{FerrerMcCowan2009} Ferrer-i Cancho, R. and McCowan, B. (2009). A law of word meaning in dolphin whistle types. Entropy, 11(4):688-701:409-419

\bibitem{Corraletal2015} Corral, Á., Boleda, G., and Ferrer-i Cancho, R. (2015). Zipf's law for word frequencies: Word forms versus lemmas in long texts. PloS one, 10(7):e0129031

\bibitem{Nabeshima_Gunji2004} Nabeshima, T. and Gunji, Y.-P. (2004). Zipf's law in phonograms and weibull distribution in ideograms: comparison of english with japanese. Biosystems, 73(2):131-139

\bibitem{baroni08} Baroni, M. (2008). Distributions in text. In Lüdeling, A. and Kytö, M., editors, Corpus linguistics: An international handbook, chapter 39. Anke Lüdeling and Merja Kytö, Berlin

\bibitem{Samlowskietal2011} Samlowski, B., Möbius, B., and Wagner, P. (2011). Comparing syllable frequencies in corpora of written and spoken language. In Proceedings of Interspeech 2011, pages 637-640

\bibitem{Farnetani_Recasens2010} Farnetani, E. and Recasens, D. (2010). Coarticulation and Connected Speech Processes, pages 316-352. Blackwell Publishing Ltd.

\bibitem{Glass2003} Glass, J. (2003). A probabilistic framework for segment-based speech recognition. Computer Speech \& Language, 17(2-3):137-152

\bibitem{TylerCutler2009} Tyler, M. D. and Cutler, A. (2009). Cross-language differences in cue use for speech segmentation. The Journal of the Acoustical Society of America, 126(1):367-376

\bibitem{Taylor2009} Taylor, P. (2009). Text-to-Speech Synthesis. Cambridge University Press. Cambridge Books Online

\bibitem{Kuhl2003}  Kuhl, P. K. (2003). Human speech and birdsong: Communication and the social brain. Proceedings of the National Academy of Sciences of the USA, 100(17):9645-9646 

\bibitem{Stegmann2013}  Stegmann, U. (2013). Animal communication theory: information and influence. Cambridge University Press

\bibitem{Doyleetal2009}  Doyle, L. R., McCowan, B., Johnston, S., and Hanser, S. F. (2011). Information theory, animal communication, and the search for extraterrestrial intelligence. Acta Astronautica, 68

\bibitem{bunge1984} Bunge, M. (1984). What is pseudoscience? The Skeptical Inquirer, 9:36-46

\bibitem{Kohler2005}  Kohler, R. (2005). Synergetic linguistics, in Quantitative linguistics, volume 760774. de Gruyter Berlin, New York

\bibitem{crystal1988segmental} Crystal, T. H. and House, A. S. (1988). Segmental durations in connected-speech signals: Current results. The Journal of the Acoustical Society of America, 83(4):1553-1573

\bibitem{luque2015scaling} Luque, J., Luque, B., and Lacasa, L. (2015). Scaling and universality in the human voice. Journal of The Royal Society Interface, 12(105):20141344

\bibitem{rodriguez2010kalaka} Rodriguez-Fuentes, L. J., Penagarikano, M., Bordel, G., Varona, A., and Diez, M. (2010). Kalaka: A tv broadcast speech database for the evaluation of language recognition systems

\bibitem{CorralJSTAT} Corral, A. (2009). Point-occurrence self-similarity in crackling-noise systems and in other complex systems. Journal of Statistical Mechanics: Theory and Experiment, 2009(01):P01022

\bibitem{brumm2005acoustic} Brumm, H. and Slabbekoorn, H. (2005). Acoustic communication in noise. Advances in the Study of Behavior, 35(35):151-209

\bibitem{Bak} Bak, P. (1996). How Nature Works. Copernicus

\bibitem{rainfall} Peters, O., Deluca, A., Corral, A., Neelin, J. D., and Holloway, C. E. (2010). Universality of rain event size distributions. Journal of Statistical Mechanics: Theory and Experiment, 2010(11):P11030

\bibitem{clauset2009power} Clauset, A., Shalizi, C. R., and Newman, M. E. (2009). Power-law distributions in empirical data. SIAM review, 51(4):661-703

\bibitem{Morenoetal2016} Moreno-Sanchez, I., Font-Clos, F., and Corral, A. (2016). Large-scale analysis of zip's law in english texts. PLoS One, 11(1):1-19 

\bibitem{zanette2014statistical} Zanette, D. H. (2014). Statistical patterns in written language. arXiv preprint arXiv:1412.3336

\bibitem{5395374} Ferrer-i Cancho, R. and Hernandez Fernandez, A. (2008). Power laws and the golden number. In Problems of general, germanic and slavic linguistics, pages 518-523. Books-XXI

\bibitem{egghe2007untangling} Egghe, L. (2007). Untangling herdan's law and heaps' law: Mathematical and informetric arguments. Journal of the American Society for Information Science and Technology, 58(5):702-709

\bibitem{lu2013deviation} L{\"u}, L., Zhang, Z.-K., and Zhou, T. (2013). Deviation of Zipf's and heaps' laws in human languages with limited dictionary sizes. Scientific reports, 3

\bibitem{Gustisonetal2016} Gustison, M. L., Semple, S., Ferrer i Cancho, R., and Bergman, T. J. (2016). Gelada vocal sequences follow menzerath's linguistic law. Proceedings of the National Academy of Sciences of the USA, 113(19):E2750-E2758 

\bibitem{greenbergetal2003}  Greenberg, S., Carvey, H., Hitchcock, L., and Chang, S. (2003). Temporal properties of spontaneous speech-a syllable-centric perspective. Journal of Phonetics, 31(3):465-485

 \bibitem{macNeilage08book} MacNeilage, P. (2008). The Origin of Speech. Oxford University Press

\bibitem{Kello2013} Kello, C. (2013). Critical branching neural networks. Psychological Review, 120

\bibitem{langton1990computation} Langton, C. G. (1990). Computation at the edge of chaos: phase transitions and emergent computation. Physica D, 42(1):12-37

\bibitem{kvale1993segmentation} Kvale, K. (1993). Segmentation and labelling of speech

\bibitem{evans2009myth-gsc}  Evans, N. and Levinson, S. (2009). The myth of language universals: Language diversity and its importance for cognitive science. Behavioral and Brain Sciences, 32(05):429-448

% \bibitem{fenk1993menzerath} Fenk, A. and Fenk-Oczlon, G. (1993). Menzerath's law and the constant flow of linguistic information. In Contributions to quantitative linguistics, pages 11-31. Springer

% \bibitem{altmann1980prolegomena} Altmann, G. (1980). Prolegomena to menzerath\'s law. Glottometrika, 2(2):1-10

% \bibitem{brehm1987description} Brehm, H. and Stammler, W. (1987). Description and generation of spherically invariant speech-model signals. Signal Processing, 12(2):119-141

% \bibitem{davenport1952experimental} Davenport Jr, W. B. (1952). An experimental study of speech-wave probability distributions. The Journal of the Acoustical Society of America, 24(4):390-399

% \bibitem{endler1993some} Endler, J. A. (1993). Some general comments on the evolution and design of animal communication systems. Philosophical Transactions of the Royal Society of London B: Biological Sciences, 340(1292):215-225

% \bibitem{gazor2003speech} Gazor, S. and Zhang, W. (2003). Speech probability distribution. IEEE Signal Processing Letters, 10(7):204-207

% \bibitem{hernandez2013ley} Hernández-Fernández, A. and Diéguez-Vide, F. (2013). La ley de Zipf y la detección de la evolución verbal en la enfermedad de alzheimer. Anuario de psicología/The UB Journal of psychology, 43(1):67-82

% \bibitem{hvrebivcek1995text} Hřebíček, L. (1995). Text levels: language constructs, constituents and the Menzerath-Altmann law. WVT, Wiss. Verlag Trier

% \bibitem{kohler1990elemente} Köhler, R. (1990). Elemente der synergetischen linguistik. Glottometrika, 12:179-187

% \bibitem{menzerath1954architektonik} Menzerath, P. (1954). Die Architektonik des deutschen Wortschatzes. Phonetische Studien. F. Dümmler

% \bibitem{misteli2009self} Misteli, T. (2009). Self-organization in the genome. Proceedings of the National Academy of Sciences of the USA, 106(17):6885-6886

% \bibitem{Paez_1972} Paez, M. and Glisson, T. (1972). Minimum mean-squared-error quantization in speech PCM and DPCM systems. IEEE Transactions on Communications, 20(2):225-230

% \bibitem{polikarpov2006cognitive} Polikarpov, A. A. (2006). Cognitive model of lexical system evolution and its verification

% \bibitem{richards1964statistical} Richards, D. (1964). Statistical properties of speech signals. In Proceedings of the Institution of Electrical Engineers, volume 111, pages 941-949. IET

% \bibitem{teupenhayn1984clause} Teupenhayn, R. and Altmann, G. (1984). Clause length and menzerath's law. Glottometrika, 6:127-138

% \bibitem{Lu2010} Lü, L., Zhang, Z.-K., and Zhou, T. (2010). Zipf's law leads to heaps' law: Analyzing their relation in finite-size systems. PLoS ONE, 5(12):e14139

% \bibitem{Menzerath1928} Menzerath, P. and de Oleza, J. M. (1928). Spanische Lautdauer. (Phonetische untersuchungen). Berlin und Leipziq : Verlag von Walter de Gruyter \& Co. 

% \bibitem{DBLP:journals/corr/Ferrer-i-Cancho15} i Cancho, R. F., Bentz, C., and Seguin, C. (2015). Compression and the origins of Zipf's law of abbreviation. arXiv:1504.0488

% \bibitem{Lachmannetal2001}  Lachmann, M., Szamado, S., and Bergstrom, C. T. (2001). Cost and conflict in animal signals and human language. Proceedings of the National Academy of Sciences of the USA, 98(23):13189-13194

% \bibitem{Hoitetal1994} Hoit, J. D., Shea, S. A., and Banzett, R. B. (1994). Speech production during mechanical ventilation in tracheostomized individuals. Journal of Speech, Language, and Hearing Research, 37(1):53-63

% \bibitem{MacLarnon2011}  MacLarnon, A. (2011). The anatomical and physiological basis of human speech production: adaptations and exaptations. Oxford University Press


\end{thebibliography}

\end{document}